\documentclass{article}
\usepackage{spconf,amsmath,graphicx}
\usepackage{breqn}
\usepackage{amsfonts}
\usepackage{soul}
\usepackage{graphicx}
\usepackage{amsmath}
\usepackage{microtype}
\usepackage{hyperref} 
\usepackage{booktabs}
\usepackage{multirow}
\usepackage{pifont}
\usepackage{float}
\usepackage{xcolor}
\usepackage{bm}
\usepackage{multicol}
\usepackage{array}
\usepackage{comment}
\usepackage{hyperref}
\usepackage{mwe} 
\usepackage{enumitem}
\setlist{nosep, leftmargin=14pt}

\title{SS-3DCapsNet: Self-supervised 3D Capsule Networks for Medical Segmentation on Less Labeled Data}
\name{Author(s) Name(s)}
\address{Author Affiliation(s)}
\name{Minh Tran$^{\star, 1}$ \qquad Loi Ly$^{\dagger}$ \qquad Binh-Son Hua $^{\ddagger}$  \qquad Ngan Le$^{\star}$}

\address{$^{\star}$ University of Arkansas \qquad
    $^{\dagger}$ Cyberlogitec Vietnam  \qquad
    $^{\ddagger}$ VinAI}

\begin{document}
%
\maketitle
\begin{abstract}
Capsule network is a recent new deep network architecture that has been applied successfully for medical image segmentation tasks. 
This work extends capsule networks for volumetric medical image segmentation with self-supervised learning. 
To improve on the problem of weight initialization compared to previous capsule networks, we leverage self-supervised learning for capsule networks pre-training, where our pretext-task is optimized by self-reconstruction. 
Our capsule network, \emph{SS-3DCapsNet}, has a UNet-based architecture with a 3D Capsule encoder and 3D CNNs decoder. Our experiments on multiple datasets including iSeg-2017, Hippocampus, and Cardiac demonstrate that our 3D capsule network with self-supervised pre-training considerably outperforms previous capsule networks and 3D-UNets.   \href{https://github.com/trqminh/ISBI2022-SS-3DCapsNet}{Code is available at here.}
\footnote[1]{Correspondence: minht@uark.edu}

\end{abstract}
\section{Introduction}
Since the introduction of UNet~\cite{ronneberger2015u,cciccek20163d}, UNet-based neural networks have achieved impressive performance in various modalities of medical image segmentation (MIS), e.g. brain tumor~\cite{brain_le_2018, le2021multi, ho2021point}, infant brain \cite{hoang2021dam, le2021offset}, liver tumor~\cite{bilic2019liver}, optic disc~\cite{ramani2020improved}, retina \cite{le2021narrow},  lung~\cite{nguyen20213ducaps}, and cell~\cite{moshkov2020test}, etc. 
Recently, capsule networks~\cite{sabour2017dynamic} have also been applied successfully for MIS~\cite{lalonde2018capsules,lalonde2021capsules,nguyen20213ducaps}.
Despite such, there remains a wide range of challenges: 
(1) Most methods are based on supervised learning, which is prone to many data problems like small-scale data, low-quality annotation, small objects, ambiguous boundaries, to name a few. 
These problems are not straightforward to overcome: labeling medical data is laborious and expensive, requiring an expert's domain knowledge.
(2) Capsule networks for medical segmentation does not outperform CNNs yet, even though the performance gap gets significantly closer~\cite{nguyen20213ducaps}.
To address such limitations and inspired by the recent success of capsule networks, in this work, we develop SS-3DCapsNet, a self-supervised capsule network for volumetric MIS. 
Our SS-3DCapsNet is built upon a state-of-the-art 3D capsule network that leverages both 3D Capsule blocks and CNN blocks for encoder and decoder architecture, respectively, which accounts for temporal relations in volumetric slices in learning contextual visual representation. 
We introduce self-supervised learning (SSL) to our 3D capsule network, which results in a UNet-like architecture that contains three pathways, i.e., visual representation, encoder, and decoder. 
The first path consists of dilated convolutional layers, which were pre-trained by SSL techniques. 
The encoder path is built upon 3D Capsule blocks, whereas the decoder path is built upon 3D CNNs blocks. 
Compared to 2D-SegCaps~\cite{lalonde2021capsules}, which is highly dependent on some random phenomena such as sampling order or weight initialization, our SS-3DCapsNet learns visual representation better as well as having a more robust weight initialization thanks to self-supervised learning.   
Compared to 3D-UCaps~\cite{nguyen20213ducaps}, we show that self-supervised learning results in additional gain in segmentation accuracy while keeping the same network complexity at test time. 

Our contributions are: (1) An effective self-supervised 3D capsules network for volumetric image segmentation. Our network architecture inherits the merits from 3D Capsule block, 3D CNN blocks, and self-supervised learning for better visual representation learning; and (2) A suite of experiments with ablation studies that empirically demonstrates the effectiveness of self-supervised 3D capsules network for MIS.

\section{Related works}
\noindent\textbf{Medical Segmentation.}
Among various DL architectures \cite{ZHOU2021102193, siddique2021u}, an encoder-decoder like UNet \cite{ronneberger2015u} and its extension have achieved impressive performance among semantic segmentation approaches. Since the seminal work of UNet \cite{cciccek20163d} for MIS, there have been numerous subsequent works in this task. 
As shown in a recent survey~\cite{lei2020medical}, MIS can be divided into two main DL groups: supervised learning and weakly supervised learning techniques.

The first group includes CNN-based supervised learning methods such as FCN \cite{long2015fully}, UNet \cite{3DUNet}, CC-3D-FCN\cite{nie20183}, RLS \cite{le2018deep}, ACRes\cite{le2021multi}, DenseVoxNet \cite{jegou2017one}, Flow-based\cite{bui2020flow},  VoxResNet \cite{chen2018voxresnet}, 3D DR-UNet \cite{vesal2018dilated}, Recurrent Level Set \cite{le2018deep}, Atrous-Net \cite{le2021multi}, Offset Curves Loss \cite{le2021offset, le2021narrow}, Point-Unet \cite{ho2021point} as notable methods. 
The second group includes weakly supervised learning methods such as transfer learning \cite{kalinin2020medical}, domain adaptation \cite{chen2019synergistic}, interactive segmentation \cite{wang2018deepigeos}. 
To address the issue of data limitation for training, Generative Adversarial Network (GAN) \cite{mirza2014conditional} has been incorporated into CNNs \cite{chang2020synthetic, le2021enhance, le2021pairflow, bui2020flow}. 
Training with imperfect datasets with scarce annotations and weak annotations has also been considered recently~\cite{tajbakhsh2020embracing}.


\noindent\textbf{Capsule Networks.} 
Capsule networks~\cite{hinton2011transforming} (CapsNet) is a new network architecture concept that strengthens feature learning by retaining more information at the aggregation layer for pose reasoning and learning the part-whole relationship, which makes it a potential solution for semantic segmentation and object detection tasks. In CapsNet, a capsule aims to represent an entity: capsule norm indicates the probability that entity is present and capsule direction indicates the configuration that entity is in. CapsNet is recently made practical \cite{sabour2017dynamic} in a CNN that incorporates two layers of capsules with dynamic routing. 

While most CapsNet has been proposed for image classification, SegCaps~\cite{lalonde2018capsules,lalonde2021capsules} expanded CapsNet for object segmentation.
This method functions by treating an MRI image as a collection of slices, each of which is then encoded and decoded by capsules to output the segmentation. 
However, SegCaps is mainly designed for 2D still images, and it performs poorly when being applied to volumetric data because of missing temporal information. 
3D-UCaps~\cite{nguyen20213ducaps} is a hybrid network architecture that utilizes both capsules and deconvolutions for feature learning and segmentation output, respectively, which shows that such combination can outperforms SegCaps design significantly in the segmentation task while retaining the merits of capsules. 
Our method further improves upon 3D-UCaps by integrating an efficient pre-training stage. 

\noindent\textbf{Self-supervised Learning.}
Self-supervised learning (SSL) is a technique for learning feature representation in a network without requiring a labeled dataset. 
A common workflow to apply SSL is to train the network in an unsupervised manner by learning with a pretext task in the pre-training stage, and then finetuning the pre-trained network on a target downstream task. 
In the case of MIS, the suitable pretext tasks can be considered in four categories: context-based, generation-based, free semantic label-based, and cross-modal-based. The first techniques utilize context features of images or videos such as context similarity \cite{caron2018deep}, spatial structure \cite{ahsan2019video}, temporal structure \cite{wei2018learning}. The second techniques have been used in image generation \cite{zhang2016colorful} and video generation \cite{srivastava2015unsupervised}. The third techniques aim to automatically generate semantic labels and applied into segmentation \cite{pathak2017learning}, contour detection \cite{pathak2017learning}. The fourth techniques are applied to multiple modalities data such as visual-audio \cite{korbar2018cooperative}, RGB-Flow \cite{sayed2018cross}.
In this work, our pretext task is based on image reconstruction. 

\section{SS-3DCapsNet: Self-supervised 3D Capsule Networks}
\label{sec:method}

We draw on the ideas of SegCaps~\cite{lalonde2018capsules} and 3D-UCaps~\cite{nguyen20213ducaps} to build our 3D capsule network for the medical segmentation task. 
Particularly, our network has three stages: (i) Visual representation, (ii) Capsule encoder, and (iii) Convolutional decoder as follows.

\begin{figure}[t]
    \centering
    \includegraphics[width=0.5\textwidth]{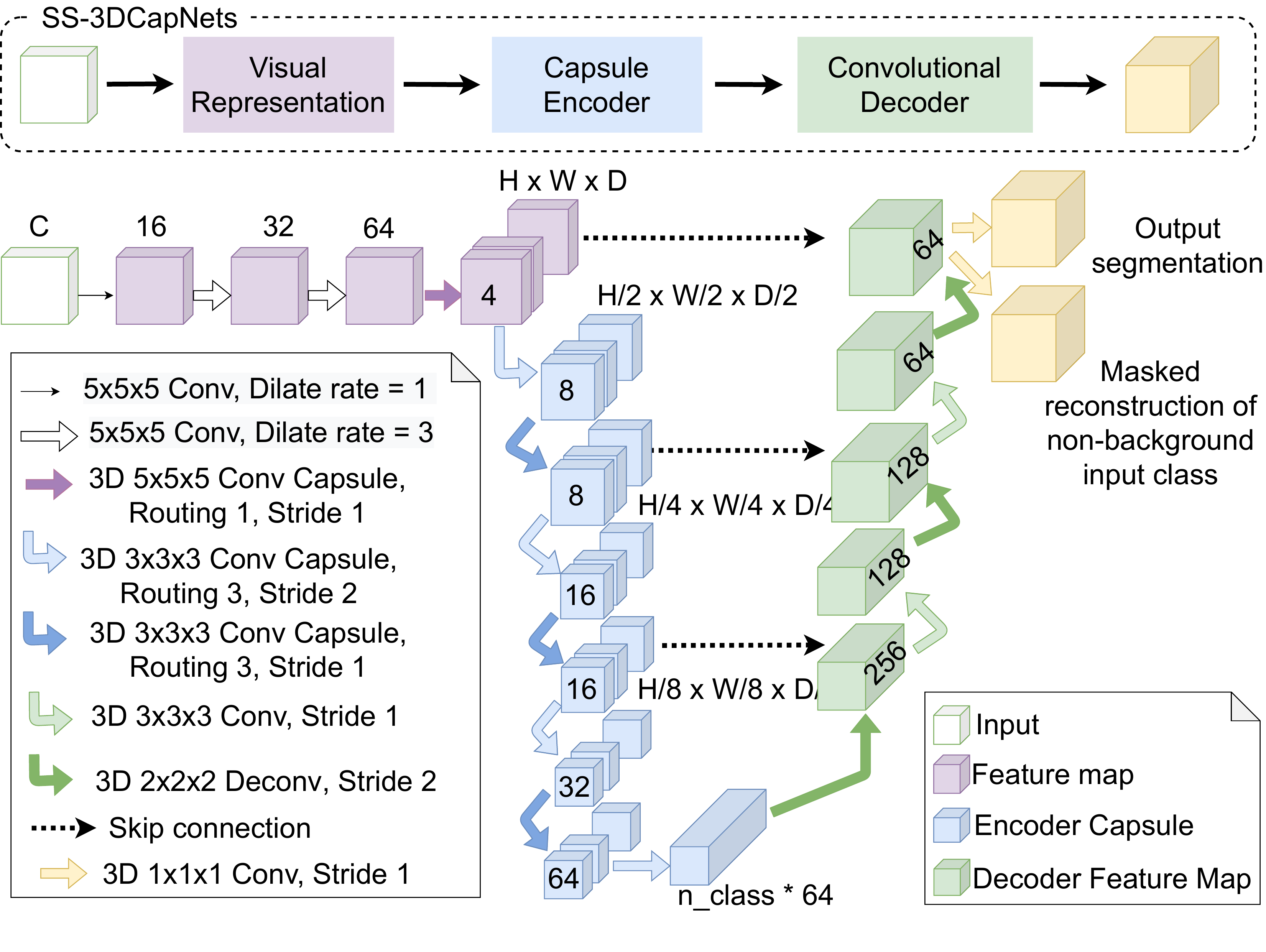}
    \caption{Our proposed SS-3DCapsNet architecture with three components: visual representation; capsule encoder, and convolution decoder. Number on the blocks indicates number of channels in convolution layer and dimension of capsules in capsule layers. \vspace{-4mm}}
    \label{fig:network}
\end{figure}
\noindent
\textbf{(i) Visual Representation:} This stage is for converting the input to a feature volume that can be consumed by the capsule encoder. Followed the concurrent work, we use three dilated convolution layers with 16, 32, 64 channels, respectively. The kernel size is set to $5 \times 5 \times 5$, with dilate rates of 1, 3, and 3, respectively. The size of the visual features is $H \times W \times D \times 64$.

\noindent
\textbf{(ii) Capsule Encoder:} In this stage, we reshape the feature volume into $H \times W \times D$ capsules, where each capsule is represented by a 64-dimensional vector. 
Here we consider both spatial and temporal data by using our 3D convolutional capsules to learn a richer representation. 
The output from a convolution capsule has the shape $H \times W \times D \times C \times A$, where $C$ is the number of capsule types and $A$ is the dimension of each capsule. 
We follow the concurrent work and set $C$ to $(16, 16, 16, 8, 8, 8)$ for each layer in the capsule encoder, respectively.
Note that as the number of capsule types in the last convolutional capsule layer is equal to the number of class labels, we can further supervise this particular layer with a margin loss~\cite{sabour2017dynamic}. 


\noindent
\textbf{(iii) Convolutional Decoder:} This is the final stage in our network. 
Here we use the decoder of 3D UNet \cite{cciccek20163d} which includes deconvolution, skip connection, convolution and BatchNorm layers \cite{ioffe2015batch} to generate the segmentation from features learned by capsule layers. Particularly, we reshape the capsules back to tensors of size $H \times W \times D \times (C \star A)$ and pass them to the decoder. The overall architecture can be seen in Fig. \ref{fig:network}.


\vspace{-5mm}
\subsection{Pretext Task}
\label{sub:SSL}
Our pretext task is self-supervised based on medical image reconstruction. 
In computer vision, it is common to use pseudo-labels defined by different image transformations, e.g, rotation, random crop, adding noise, blurring, scaling, flipping, jigsaw puzzle, etc. to supervise the pretext task.  
While such transformations work well for classification as a downstream task, since our downstream task is segmentation, we propose to use a pretext task that can consider reconstructing the original image. 
As medical images are captured in low contrast and the object-of-interest in medical images usually follows some specific patterns, we select contrast transformations to perform the pretext task with the reconstruction loss. 

\begin{figure}[t]
    \centering
    \includegraphics[width=0.85\linewidth]{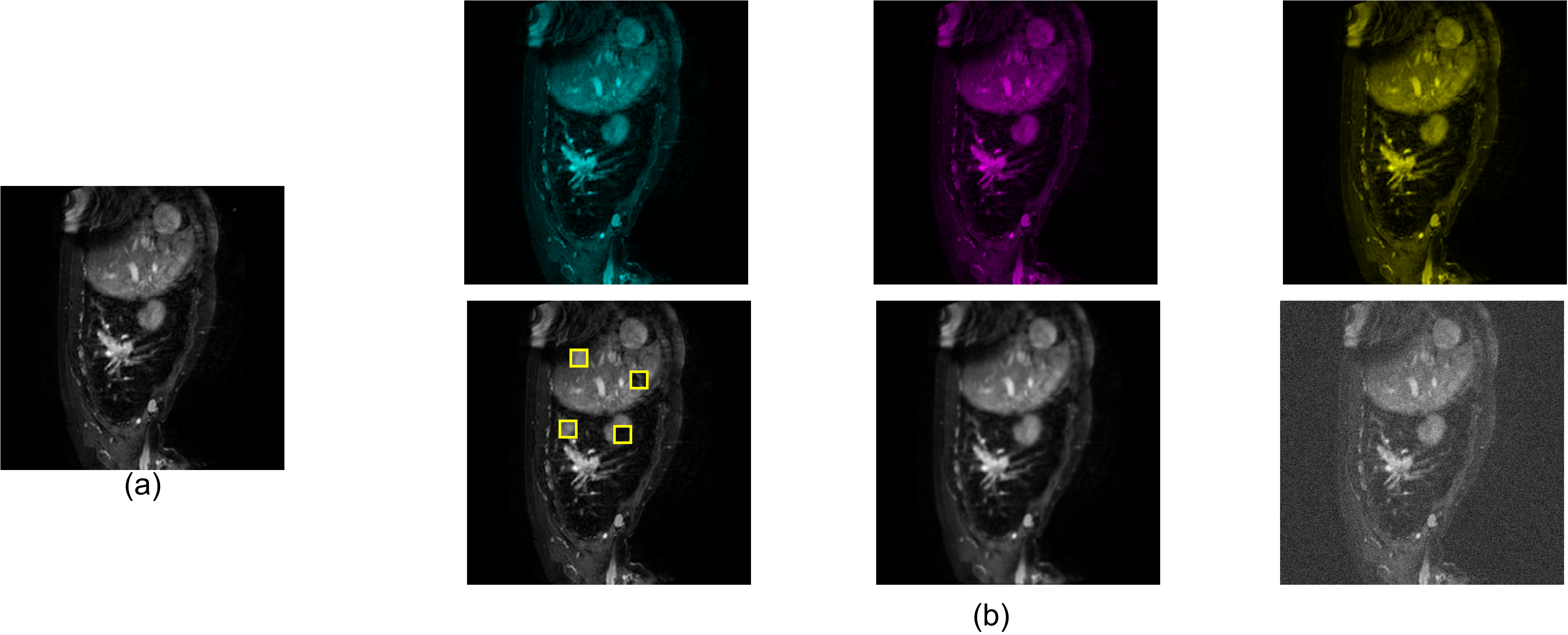}
    \caption{Examples of six transformations for self-supervised learning. (a): original image. (b) from left to right, top to bottom: zeros-green-channel, zeros-red-channel, zeros-blue-channel, swapping (4 swapped patches are shown in yellow boxes), blurring, noisy.}
    
    \label{fig:transformation}
\end{figure}

\begin{figure}[t]
    \centering
    \includegraphics[width=0.45\textwidth]{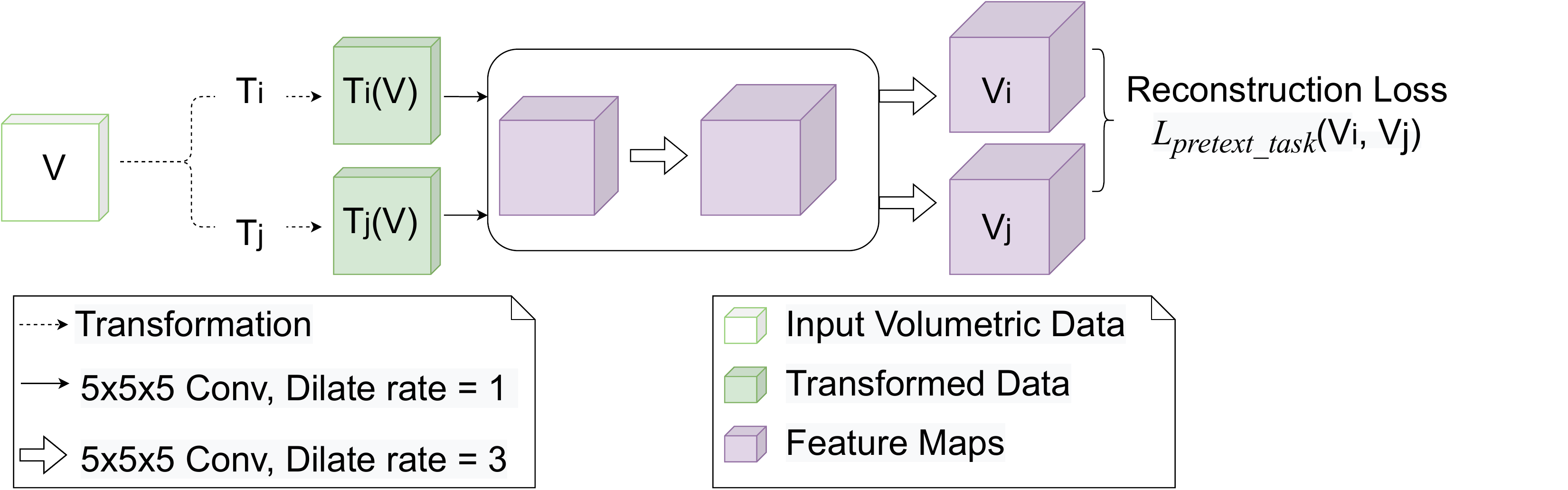}
    \caption{Our pretext task with reconstruction loss. \vspace{-5mm}}
    \label{fig:SSL}\
\end{figure}

The details of our pre-training are as follows. 
Our pretext task is based on reconstruction from various transformations i.e. noisy, blurring, zero-channels (R,G,B), swapping as shown in Fig. \ref{fig:transformation}. 
Let $\mathcal{F}$ is the visual representation network. The transformation is defined as $\{T_i\}_{i=1}^{i=N}$, where $T_0$ is an identity transformation and $N$ is set as 6 corresponding to six transformations (Fig. \ref{fig:transformation}). 
Let $V$ denote as the original input volumetric data. Our pretext task is performed by applying two random transformations $T_i, T_j (i, j \in [0, 6])$ into $V$. The transformed data is then $T_i(V)$ and $T_j(V)$, respectively. The visual feature of transformed data after applying the network $\mathcal{F}$ is $V_j$ and $V_j$, where $V_i = \mathcal{F}(T_i(V))$ and $V_j = \mathcal{F}(T_j(V))$. 
The network $\mathcal{F}$ is trained with a reconstruction loss defined by:
\begin{equation}
    \mathcal{L}_{pretext}(V_i, V_j) = ||V_i - V_j||_{2}.
\label{eq:loss}
\end{equation}
The pretext task procedure is illustrated in Fig. \ref{fig:SSL}. Note that the trivial case can occur when training this pretext task. More concretely, the $\mathcal{F}$ function can map any volumetric input data into a constant (e.g., 0). Therefore, the pre-trained model from the pretext task may have no impact on the downstream ones. Our pretext task training is probably stuck at its first local minima so that, luckily, the trivial case cannot occur. Unfortunately, in our accepted version at ISBI2022, we overlooked this issue, so we want to clarify it in this version.

\vspace{-2mm}
\subsection{Downstream Task} 
After pre-training, we train our SS-3DCapsNet network with annotated data on the medical segmentation task. The total loss function to train this downstream task is a sum of three losses:
\begin{equation}
    \mathcal{L}_{downstream} = \mathcal{L}_{margin} + \mathcal{L}_{CE} + \mathcal{L}_{reconstruction}.
\end{equation}
The margin loss is adopted from ~\cite{sabour2017dynamic} and it is defined between the predicted label $y$ and the ground truth label $y^*$ as follows:
\begin{align}
\mathcal{L}_{margin} =& y^* \times (\max(0, 0.9 - y))^2 + \\ & 0.5 \times (1 - y^*) \times (\max(0, y - 0.1))^2. \nonumber
\end{align}
Particularly, we compute the margin loss ($\mathcal{L}_{margin}$) on the capsule encoder output with downsampled ground truth segmentation. 
We compute the weighted cross-entropy loss ($\mathcal{L}_{CE}$) on the convolutional decoder.  
We also regularize the training with a network branch that aims at reconstructing the original input with masked mean-squared errors  ($\mathcal{L}_{reconstruction}$)~\cite{sabour2017dynamic,lalonde2018capsules}.

\section{Experimental Results}
\noindent
\textbf{4.1. Implementation Details}
We conduct our experiments and comparisons on iSeg \cite{wang2019benchmark}, Hippocampus, and Cardiac~\cite{simpson2019large} datasets.
For iSeg, we follow 3D-SkipDenseSeg \cite{bui2019skip} to have the training set of 9 subjects and testing set of subject \#9. 
On Hippocampus, and Cardiac~\cite{simpson2019large}, the experiments are conducted by 4-fold cross-validation.  

\begin{table}[!h]
\small
\centering
\caption{Comparison on iSeg-2017. $1^{st}$ group: 3D CNN-based networks and $2^{nd}$ group: Capsule-based networks.}
\vspace*{0.2cm}
\resizebox{\linewidth}{!}{
\label{table:iseg}
\begin{tabular}{@{}l|ll  llll@{}}
\toprule
& \multicolumn{1}{c}{\multirow{2}{*}{Method}} & \multicolumn{1}{c}{\multirow{2}{*}{Depth}} & \multicolumn{4}{c}{Dice Score}                                      \\  \cmidrule(l){4-7} 
& \multicolumn{1}{c}{}                        &             & \multicolumn{1}{c}{WM}             & \multicolumn{1}{c}{GM}             & \multicolumn{1}{c}{CSF} & Average        \\ \midrule
& Qamar et al. \cite{qamar2020variant}    & 82     & 90.50    & \textbf{92.05}   & \textbf{95.80}  & \textbf{92.77} \\
\multirow{6}{*}{\rotatebox{90}{\shortstack{CNN}}}&  3D-SkipDenseSeg \cite{bui2019skip}    & 47    & \textbf{91.02} & 91.64 & 94.88          & 92.51 \\
& VoxResNet \cite{chen2018voxresnet}                                   & 25    & 89.87                              & 90.64                              & 94.28                   & 91.60          \\
& 3D-UNet \cite{cciccek20163d}                                     & 18   & 89.83                              & 90.55                              & 94.39                   & 91.59          \\
& CC-3D-FCN \cite{nie20183}                                   & 34     & 89.19          & 90.74          & 92.40                   & 90.79          \\
& DenseVoxNet \cite{jegou2017one}                                 & 32      & 85.46          & 88.51          & 91.26                   & 89.24          \\ \midrule
\multirow{4}{*}{\rotatebox{90}{\shortstack{Capsule}}} 
& 2D SegCaps \cite{lalonde2018capsules}                                  & 16      & 82.80                              & 84.19                              & 90.19                   & 85.73          \\
& 3D-SegCaps \cite{nguyen20213ducaps}     & 16   & 86.49                              & 88.53                              & 93.62                   & 89.55  \\
& 3D-UCaps \cite{nguyen20213ducaps}  & 17 & 90.21 & 91.12 & \textbf{94.93} & 92.08\\
& \textbf{Our SS-3DCapsNet }                      & 17     & \textbf{90.78}                             & \textbf{91.48}                             & 94.92                   &\textbf{92.39}         \\ \bottomrule
\end{tabular}
}
\end{table}

We implemented our method in Pytorch. We used patch size of $64 \times 64 \times 64$ for iSeg and Hippocampus whereas patch size of $128 \times 128 \times 128$ on Cardiac. Our SS-3DCapsNet was trained without any data augmentation.
We used Adam optimizer with an initial learning rate of 0.0001. 
The learning rate is decayed by 0.05 if the Dice score on the validation set does not increase for 50,000 iterations. 
Early stopping is set at 250,000 iterations as in \cite{lalonde2018capsules}. 



\begin{table}[]
\centering
\caption{Comparison on Cardiac with 4-fold cross validation.}
\vspace*{0.2cm}
\resizebox{\linewidth}{!}{
\label{table:cardiac}
\centering
\begin{tabular}{ll|ll}
\toprule
\multicolumn{2}{l|}{3D CNN-based networks} & \multicolumn{2}{l}{Capsule-based networks} \\ \cmidrule(l){1-4} 
3D UNet\cite{cciccek20163d}  &       84.30               & SegCaps (2D)~\cite{lalonde2018capsules}   &  66.96 \\ 
3D Vnet\cite{VNet} &          84.20            & Multi-SegCaps (2D)~\cite{survarachakan2020capsule}  & 66.96 \\
3D DR-UNet \cite{vesal2018dilated}   &     87.40    & 3D-UCaps \cite{nguyen20213ducaps} & 89.69 \\ 

& &  \textbf{Our SS-3DCapsNet }  &       \textbf{89.77   }          \\  
\bottomrule
\end{tabular}
}
\end{table}

\begin{table}[!t]
\centering
\caption{Comparison on Hippocampus with 4-fold.}
\label{table:hippocampus}
\vspace*{0.2cm}
\resizebox{1.0\linewidth}{!}{
\begin{tabular}{@{}llll|lll@{}}
\toprule
\multicolumn{1}{c}{Method} & Anterior &           &        & Posterior &           &        \\ \cmidrule(l){2-7} 
\multicolumn{1}{c}{}       & Recall   & Precision & Dice   & Recall    & Precision & Dice   \\ \midrule
Multi-SegCaps (2D)~\cite{survarachakan2020capsule}               & 80.76   & 65.65    & 72.42 & 84.46    & 60.49    & 70.49 \\
EM-SegCaps (2D)~\cite{survarachakan2020capsule}                  & 17.51   & 20.01    & 18.67 & 19.00    & 34.55    & 24.52 \\
3D-UCaps \cite{nguyen20213ducaps}                       & 81.70  & 80.19  &80.99 &  80.2 & 79.25 &79.48  \\ 

\textbf{ Our SS-3DCapsNet}                       & \textbf{81.84}  &  \textbf{81.49 }  & \textbf{81.59} &  \textbf{80.71 } & \textbf{80.21 }   & \textbf{79.97} \\ \bottomrule
\end{tabular}
}
\end{table}

\begin{table}[t]
\small
\caption{Performance of SS-3DCapsNet on iSeg with different network configurations.}
\label{table:ablation}
\centering
\vspace*{0.2cm}
\resizebox{\linewidth}{!}{
\begin{tabular}{lllll}
\toprule
\multicolumn{1}{c}{\multirow{2}{*}{Method}} & \multicolumn{4}{c}{Dice Score}                                                            \\ \cline{2-5} 
\multicolumn{1}{c}{}                        & \multicolumn{1}{c}{WM}    & \multicolumn{1}{c}{GM}    & \multicolumn{1}{c}{CSF} & Average \\ \hline
change number of capsule (set to 4)    & \multicolumn{1}{c}{89.02} & \multicolumn{1}{c}{89.78} & 89.95                   & 89.58   \\
w/o visual representation           & 89.15                     & 89.66                     & 90.82                   & 89.88   \\
w/o margin loss                  & 87.62                     & 88.85                     & 92.06                   & 89.51   \\
w/o reconstruction loss          & 88.50                     & 88.96                     & 90.18                   & 89.22   \\
w/o pretext task                 & 90.21                     & 91.12                     & \textbf{94.93}                   & 92.08  \\ 
\textbf{SS-3DCapsNet}            & \textbf{90.78}                     & \textbf{91.48}                     & 94.92                   & \textbf{92.39} \\ \bottomrule
\end{tabular}
}
\end{table}

\begin{table}[!t]
\centering
\caption{Performance of SS-3DCapsNet on Precision (Pre), Recall (Rec) and Dice score (DSC) with and without pretext task on various datasets.}
\vspace*{0.2cm}
\resizebox{\linewidth}{!}{
\begin{tabular}{l|lll|lll|lll }
\toprule
           & \multicolumn{3}{c|}{iSeg} & \multicolumn{3}{c|}{Hippocamus} & \multicolumn{3}{c}{Cardiac} \\ \cmidrule(l){2-10} 
            & Pre & Rec & DSC & Pre   & Rec   & DSC   & Pre  & Rec  & DSC  \\ \midrule
w/o. SSL &     92.28  &    91.29   &   92.08   &    79.72        &  80.95  & 80.24 &       84.60    &  \textbf{95.06}      &   89.69   \\
w. SSL    &      \textbf{92.54}     &    \textbf{92.37}    &   \textbf{92.39}   &      \textbf{80.85 }     &    \textbf{81.27}    &  \textbf{80.78}    &   \textbf{86.24}    & 94.21  &  \textbf{89.77}  \\  \bottomrule
\end{tabular}
}
\label{tab:SSL}
\end{table}

\noindent
\textbf{4.2. Performance and Comparison}
We compare our SS-3DCapsNet with both SOTA 3D CNNs-based and Capsule-based segmentation methods. 3D-Ucaps \cite{nguyen20213ducaps} has two versions of with and without utilizing MONAI \cite{MonAI}. To conduct a fair comparison, we report the version without MONAI.


The comparison between our proposed SS-3DCapsNet with SOTA segmentation approaches on iSeg dataset \cite{wang2019benchmark} is given in Table~\ref{table:iseg}. 
As can be seen, 3D capsule networks (3D-SegCaps, SS-3DCapsNet) outperform 2D-SegCaps by a wide margin. 
This performance gap can be explained by the combination of pre-training, Capsule encoder, and Convolutional decoder in SS-3DCapsNet. 
Our SS-3DCapsNet also outperforms 3D-SegCaps, which contains only a Capsule-based encoder and decoder. 
Our SS-3DCapsNet also performs comparably to SOTA 3D CNNs, but our network is significantly shallower (17 layers vs. 82 layers in \cite{qamar2020variant}). 
Our network also has fewer parameters and a better Dice score when compared to  SOTA 3D CNNs with similar number of layers, e.g. 3D-UNets \cite{cciccek20163d} (18 layers). In addition to iSeg, we also evaluate our SS-3DCapsNet on Hippocampus and Cardiac, where the results are shown in Table \ref{table:cardiac} and Table \ref{table:hippocampus}.

\noindent
\textbf{4.3. Ablation Study}
We analyze the performance of our method as follows.

\textbf{i. Network Configuration:} We trained SS-3DCapsNet under various settings as shown in Table~\ref{table:ablation}. 
By following the concurrent work on 3D capsule networks, we use a baseline where the number of capsules of the first layer is reduced to 4 (similar to SegCaps). 
As can be seen, each component including visual representation, margin loss, reconstruction loss, pre-training contributes to the final performance. Removing any of such components would result in performance drops.

\textbf{ii. SSL Contribution}: We perform experiments on various datasets and turn on/off the self-supervision step in the experiments. The results in Table \ref{tab:SSL} clearly shows that pre-training plays an important role in our method, which improves the Dice score considerably in iSeg, and slightly in other datasets.

\vspace{3mm}
\noindent
\textbf{CONCLUSION}

In this work, we proposed a capsule network for MIS powered with self-supervised pre-training. Our SS-3DCapsNet can both utilize self-supervised learning and 3D capsules for learning features while retaining the advantage of traditional convolutions in decoding the segmentation results. Even though we use capsules with dynamic routing only in the encoder of a simple Unet-like architecture, we can achieve the competitive result with the SOTA models on iSeg-2017 challenge while outperforming SegCaps~\cite{lalonde2018capsules} on different complex datasets with less labeled annotated data. 
Future work includes exploring different self-supervised learning methods such as SimCLR~\cite{chen2020simple} for better feature learning and representation.

\vspace{3mm}
\noindent
\textbf{Acknowledgment}
This material is based upon work supported by the National Science Foundation under Award No. OIA-1946391 and NSF Track-2 WVAR-CRESH. 
\scriptsize
\bibliographystyle{ieeetr}
\setlength{\parskip}{-1pt}
\bibliography{refs}

\end{document}